\documentclass[12pt]{article}
\usepackage{amsmath,amsfonts, amssymb, braket, bbold}
\usepackage{tikz}
\usetikzlibrary{trees,er,snakes,shapes,mindmap}
\usepackage{titlesec}
\titleformat{\section}
 {\normalfont\fontfamily{put}\fontsize{12pt}{16pt}\bfseries\color{black}}
{\thesection}{1em}{}
\titleformat{\subsection}
 {\normalfont\fontfamily{put}\fontsize{12pt}{16pt}\bfseries\color{black}}
{\thesubsection}{1em}{}
\def \beq  {\begin{equation}}
\def \eeq  {\end{equation}}
\def \beqar {\begin{eqnarray}}
\def \eeqar {\end{eqnarray}}
\allowdisplaybreaks
\def\sqr#1#2{{\vcenter{\vbox{\hrule height.#2pt
\hbox{\vrule width.#2pt height#1pt \kern#1pt
\vrule width.#2pt}\hrule height.#2pt}}}}

\def\S {{\cal S}}
\def\la {{\langle}}
\def\ra {{\rangle}}
\def\vx {{\vec x}}
\def\vy {{\vec y}}
\def\vk {{\vec k}}
\def\vf {{\varphi}}

\def\dag {{\dagger}}

\def\vq {{\vec q}}

\def\Tr {{\rm Tr}}

\def\ba {\bar{a}}

\def\bA {\bar{A}}

\def\bi{\bar{i}}

\def\vk {\vec{k}}
\def\vx {{\vec x}}
\def\vz {\vec{z}}
\def\vy{\vec{y}}

\def\vw {\vec{w}}

\def\bnabla {\bar{\nabla}}

\def\dag {\dagger}
\def\del {\partial}
\def\bdel{\bar{\partial}}
\def\a {\alpha}

\def\e {\epsilon}
\def\d {\delta}

\def\bz {{\bar{z}}}

\def\A {{\cal A}}

\def\D {{\cal D}}
\def\E {{\cal E}}

\def\H {{\cal H}}

\def\vf {{\varphi}}

\def\half{\textstyle{1\over 2}}

\mathchardef\mhyphen="2D
\begin{document}
\fontfamily{bch}\fontsize{12pt}{16pt}\selectfont
\def \CMP {{Commun. Math. Phys.}}
\def \PRL {{Phys. Rev. Lett.}}
\def \PL {{Phys. Lett.}}
\def \NPBProc {{Nucl. Phys. B (Proc. Suppl.)}}
\def \NP {{Nucl. Phys.}}
\def \RMP {{Rev. Mod. Phys.}}
\def \JGP {{J. Geom. Phys.}}
\def \CQG {{Class. Quant. Grav.}}
\def \MPL {{Mod. Phys. Lett.}}
\def \IJMP {{ Int. J. Mod. Phys.}}
\def \JHEP {{JHEP}}
\def \PR {{Phys. Rev.}}
\def \JMP {{J. Math. Phys.}}
\def \GRG{{Gen. Rel. Grav.}}
\begin{titlepage}
\null\vspace{-62pt} \pagestyle{empty}
\begin{center}
\vspace{1.3truein} {\large\bfseries
On Casimir effect in Yang-Mills theories in three and}
\vskip .1in
{\large\bfseries four dimensions}\\
~\\
~\\
{\sc Dimitra Karabali$^{a,c}$, Antonina Maj$^{c}$, V.P. Nair$^{b, c}$}\\
\vskip .2in
{\sl $^a$Physics and Astronomy Department,
Lehman College, CUNY\\
Bronx, NY 10468}\\
\vskip.1in
{\sl $^b$Physics Department,
City College of New York, CUNY\\
New York, NY 10031}\\
\vskip.1in
{\sl $^c$The Graduate Center, CUNY\\
New York, NY 10016}\\
 \vskip .1in
\begin{tabular}{r l}
{\sl E-mail}:&\!\!\!{\fontfamily{cmtt}\fontsize{11pt}{15pt}\selectfont 
dimitra.karabali@lehman.cuny.edu}\\
&\!\!\!{\fontfamily{cmtt}\fontsize{11pt}{15pt}\selectfont amaj@gradcenter.cuny.edu}\\
&\!\!\!{\fontfamily{cmtt}\fontsize{11pt}{15pt}\selectfont vpnair@ccny.cuny.edu}\\
\end{tabular}
\vskip .5in

\centerline{\large\bf Abstract}
\end{center}
We calculate the Casimir energy for the configuration of
two parallel plates coupled to
nonabelian gauge fields with a Yang-Mills action. We consider both
2+1 and 3+1 dimensions in the manifestly gauge-invariant formalism
we have pursued over the last several years which allows us to factor 
out the gauge degrees of freedom. A boundary action in the functional integral,
equivalent to the insertion of operators representing the plates,
is used to enforce the required boundary conditions
for the gauge fields. The result is for a kinematic regime corresponding to the exchange of gluons with a dynamically generated mass.  We find good agreement in 2+1 dimensions and reasonable agreement
in 3+1 dimensions with lattice-based numerical evaluations.

\end{titlepage}
\fontfamily{bch}\fontsize{12pt}{16pt}\selectfont
\pagestyle{plain} \setcounter{page}{2}
\section{Introduction}
Casimir effect has recently emerged as an interesting probe of certain nonperturbative aspects of nonabelian gauge theories in both three and four dimensions.
There have been several lattice-based numerical simulations
of geometrical set-ups
including the classic parallel plate arrangement  
(or parallel wires in three dimensions)
\cite{{chern1},{chern2},{ngwenya}}.
The results on the Casimir energy, say, between the two
parallel plates, have been quite revealing.
In the perturbative regime where
the propagating particles can be taken to be massless gluons, we should
expect the familiar result for massless fields with
the power law fall-off of the Casimir energy as a function of the separation of the plates. On the other hand, in the confining regime, the only correlation
between the plates must involve the exchange of massive colorless glueballs
and hence we should expect exponential fall-off with the exponent related to the lowest glueball mass.
The actual simulation yields a result which agrees with neither of these cases.
There is exponential fall-off, unlike the massless case, but the exponent is much smaller than what is expected from the glueball exchange.
In fact, simulations seem to capture an intermediate regime where the gluon
should be treated as a massive particle, with the nonperturbative
dynamical generation of mass, but can still be considered as a propagating degree of freedom.

The natural question then is whether we can understand the dynamical
generation of mass, starting from first principles and then use that result to calculate the Casimir energy.
Here we will argue that this is indeed possible.
In fact, in the case of (2+1)-dimensional Yang-Mills theory, 
a nonperturbative Hamiltonian analysis led to a prediction for the propagator
mass of the gluon \cite{{KKN1}, {KNY}, {Ham2+1}}.
 The resulting expression for the Casimir energy
was shown to agree remarkably well with the lattice-based numerical
results \cite{KN1}.
While our Hamiltonian analysis applies to all gauge groups, 
at the time of reference \cite{KN1}, the lattice calculation of the
Casimir energy was available only for the case of $SU(2)$ Yang-Mills
theory \cite{chern1}, and so this was the case for which the comparison was made.
Recently, there have been
estimates of the Casimir energy for the case of $SU(3)$
\cite{ngwenya}.
In the following, we will therefore first present the comparison of these
results with our Hamiltonian analysis.
We note that the authors of \cite{{chern1},{ngwenya}} used a fitting formula for
their results, which is different from ours.
The mass defined by the exponential fall-off was interpreted
as due to a state bound to the plates.
The interpretation we give here in terms of exchange of
gluons with dynamically generated mass is more natural and also supported by the Hamiltonian analysis.
The agreement of our formula with the data is very good as well.

A further point is that the mass parameter is related to the magnetic mass
of the (3+1)-dimensional theory in the high temperature limit.
The argument is that at high temperatures the (3+1)-dimensional theory
will reduce to a 3d Euclidean theory, which, via the usual Wick rotation
can be related to the Hamiltonian analysis we have done.
The propagator mass then controls the exponential fall-off
in the correlators of the 3d Euclidean theory, which in turn implies that it
is the magnetic screening mass of the (3+1)-dimensional theory
\cite{{GPY},{magmass}}.
It is important that the value of the mass obtained in the Hamiltonian
analysis \cite{Ham2+1} simultaneously gives good agreement for the Casimir energy \cite{KN1} and
also for estimates of the magnetic screening mass
by other methods \cite{magmass}.

Turning to 3+1 dimensions, we note that an analysis similar to what was done
for 2+1 dimensions is possible.
The idea is to consider the complex Euclidean version
of the spacetime manifold and view it as a complex manifold.
A simple choice is then the manifold $\mathbb{CP}^2$, where
we are able to obtain a parametrization of gauge potentials
similar to what was used in the Hamiltonian analysis in 2+1 dimensions
\cite{KMN}.
The complex structure is crucial for this and the parametrization allows for
factoring out the gauge degrees of freedom without having to use
gauge-fixing.
The gauge-invariant volume element was calculated and shown 
to require the specification of a dynamical mass for the gluons.
The 4-dimensional theory, as is well-known, requires dimensional
transmutation: one needs to specify an input parameter with the dimensions 
of mass to define the theory. Conventionally, this is taken to be
$\Lambda_{\rm QCD}$, defined via one-loop running of the coupling
constant. However, in our formulation, the dynamical mass of the gluons
serves as the input parameter. In principle, $\Lambda_{\rm QCD}$ is then calculable
in terms of this dynamical mass, if one performs perturbative
calculations. We have not done this yet. But the key point is that
our analysis justifies calculating the Casimir energy in terms of
a massive field with two polarizations. The resulting formula, we will show,
is in reasonable agreement with the available lattice data
from \cite{ngwenya} for $SU(2)$ and from
\cite{{chern2},{ngwenya}} for $SU(3)$.

There are several approaches to the computation of the Casimir energy.
The most common method is to calculate the vacuum energy with the field modes restricted to obey appropriate boundary conditions at the locations of the plates \cite{casimir}. The difference with the true vacuum, which does not have
any such boundary conditions for the modes, is the Casimir energy.
An alternative is to view the plates as physical entities and calculate
their interaction energy, via van der Waals type forces.
Yet another approach is to impose the boundary conditions by
constraints and obtain the partition function and free energy.
This last method has the advantage of bypassing normalization issues
in the functional integral, dealing directly with correlations between the
plates. This is what we shall use in what follows.

In section 2, we give the general method of calculating the Casimir energy
in terms of the constraints as explained above, for both scalar fields
and for gauge fields. The specific calculation for 2+1 dimensions is given
in section 3 and for 3+1 dimensions in section 4. We also discuss the dependence of the Casimir energy on different boundary conditions. The comparison of these results with the lattice-based numerical evaluation is done in section 5,
with a number of relevant comments. 
We give a resum\'e of our conclusions in section 6.
The paper includes a short Appendix on how the results of our analysis on $\mathbb{CP}^2$ can be expressed in
a form which has 4d-covariance, suitable for application
to the calculation of the Casimir energy.

\section{Constraints and boundary actions}

In this section we will frame the calculation of the Casimir energy
by imposing boundary conditions as constraints in the functional integral.
The essence of the calculation can then be reduced to that of a
two-point function for the gauge potentials.
To see how this works out in general, we will start with a massless
scalar field described by the Euclidean action
\beq
\S =  \int d^4x\, {1\over 2} (\del \phi )^2
\label{cas1}
\eeq
We will consider the parallel plate geometry where we impose
the boundary condition that $\phi$ vanishes on the plates.
(As is well-known, the self-adjointness of the Laplacian allows for an infinity of boundary conditions. Boundary actions suitable for the general case in the context of Casimir energy have been discussed in \cite{KN2},
see also \cite{asorey}.
But here we will focus on fields vanishing on the plates.)
In turn, this requires expanding in terms of Dirichlet modes
and then calculating the determinant from the functional integral.
However, we can also impose the boundary condition by a functional 
$\delta$-function on the plates.
Thus we consider the functional integral
\beq
Z = \int [D\phi][ D\sigma] \, e^{ - S} \, e^{i \int \sigma \phi }
\label{cas2}
\eeq
We take the plates to be at $x_3 = 0, \, R$ and the field
$\sigma$ is a 3-dimensional field on the plates. Actually, we have
two independent fields $\sigma_1 = \sigma(\vx, x_3 = 0)$ and
$\sigma_2 = \sigma(\vx, x_3 = R)$.
Thus 
\beq
i \int \sigma \phi = i \int d^3x\, \sigma_1(\vx) \phi(\vx, 0)
+ i \int d^3x\, \sigma_2 (\vx ) \phi(\vx, R)
\label{cas3}
\eeq
Integration over $\sigma$ gives the $\delta$-function setting
$\phi = 0$ on the plates.
As for the integration over the $\phi$-field, there are no boundary conditions to be imposed at the location of the plates. So we can integrate it out to get
a reduced action for the $\sigma$-fields and then do the integration over them.
The propagator for $\phi$ has the usual form
\beq
\la \phi (x) \phi(y)\ra = \int {d^4k \over (2\pi)^4}\, {e^{ -i k (x-y)} \over 
\omega^2 + k_3^2}
\label{cas4}
\eeq
where $\omega^2 = k_1^2 + k_2^2 + k_4^2$. Integration over the $\phi$'s will
require
\beqar
K_{00}(\vx, \vx')=K_{\rm RR}(\vx, \vx')&=&\la \phi(\vx, 0) \phi(\vx', 0) \ra = \int {d^3 k \over (2\pi)^3} 
{1\over 2 \omega} e^{ -i \vk\cdot(\vx - \vx')}\nonumber\\
K_{\rm  0R}(\vx, \vx') =K_{\rm R0}(\vx, \vx')&=& \la \phi(\vx, 0) \phi(\vx', R) \ra = \int {d^3 k \over (2\pi)^3} 
{1\over 2 \omega} e^{ -\omega  R}\, e^{ -i \vk\cdot(\vx - \vx')}
\label{cas5}
\eeqar
Integrating out $\phi$, we get
\beq
Z = \bigl(\det{(-\del^2/2\pi)}\bigr)^{-{\half}}~
\int [D\sigma] \, \exp\left( - {1\over 2} \int \sigma_1 K_{00} \sigma_1
 - {1\over 2} \int \sigma_2 K_{\rm RR} \sigma_2 - \int \sigma_1 K_{\rm 0R} \sigma_2\right)
 \label{cas6}
\eeq
The determinant is evaluated with modes over all of space, so
it does not depend on $R$. Hence it is irrelevant for the Casimir energy.
Consider now integrating out $\sigma_1$. This will give another determinant
(of $K_{00}$), which is also independent of $R$. The result for the
partition function then takes the form
\beqar
Z &=& {\cal N} \, \int [D\sigma_2] \exp\left( 
- {1\over 2} \int \sigma_2 \bigl(K_{\rm RR} - K_{\rm R0} K^{-1}_{00} K_{\rm 0R} \Bigr) \sigma_2
\right)\nonumber\\
&=& {\cal N}' \, \left[ \det\bigl( K_{\rm RR} - K_{\rm R0} K^{-1}_{00} K_{\rm 0R} \bigr) \right]^{-\half}
\label{cas7}
\eeqar
The normalization factors are independent of $R$.
The $R$-dependent part of the partition function is thus given by
\beq
-\log Z =  {1\over 2} \Tr \log \Bigl( 1 - K^{-1}_{\rm RR} K_{\rm R0} K^{-1}_{00} K_{\rm 0R} \Bigr)
\label{cas8}
\eeq
Notice that $K^{-1}_{\rm RR}$ and $K^{-1}_{00} $ are separately
on the plates
at $x_3 = R$ and $x_3 = 0$ respectively, so they do not involve
the separation $R$ and we find
\beq
K^{-1}_{00} = K^{-1}_{\rm RR} = \int {d^3k \over (2\pi)^3} (2 \omega) \, e^{-i \vk\cdot(\vx- \vx')}
\label{cas9}
\eeq
It is then easy to see that the $R$-dependent part of the
partition function takes the form
\beqar
-\log Z &=& {1\over 2} \int d^3x \int {d^3k \over (2\pi)^3}
\log (1- e^{- 2 \omega R})\nonumber\\
&=& - {1 \over 4 \pi^2} \int d^3x \sum\int_0^\infty d\omega \omega^2 
{e^{- 2 \omega R n} \over n}\nonumber\\
&=& - {1\over 16 \pi^2} {1\over R^3} \zeta(4) \int d^3x
 = - {\pi^2 \over 1440\, R^3}\, \A \, T
\label{cas10}
\eeqar
The integral of $d^3x$ gives the expected area factor $\A$ and the length of the time-direction $T$. The latter gets removed in the free energy,
so that the Casimir energy per unit area is given as
\beq
{\E\over \A} = - {\pi^2 \over 1440\, R^3}
\label{cas10a}
\eeq
This is exactly the answer we get if we consider just the free action 
but use a mode expansion for the fields $\phi$
which obey
Dirichlet (vanishing) conditions on the plates.
The advantage of this calculational procedure
 is that the correlation functions in (\ref{cas5})
are evaluated in free space with no plates.

It is well-known that the partition function in the presence of a
heavy charged particle is given by the functional integral of $e^{-\S}$
with the insertion of a Wilson line. The two terms
$e^{i \int \sigma_1 \phi}$ and $e^{i \int \sigma_2 \phi}$ can be viewed as
the analogs, corresponding to the insertion of the plates.
When we consider the plates coupled to
a gauge field, we need the corresponding boundary action
which captures the required boundary conditions.
This action is of the form
\beq
S_b = 2\,  i \int  C^a \, F^a  = {i }\int \begin{cases}
 C^a  \e^{ij} F^a_{ij}&\hskip .2in {2+1~ {\rm dim}}\\
C_\mu^a  \e^{\mu\nu\alpha} F^a_{\nu\alpha}&\hskip .2in {3+1~ {\rm dim}}\\
\end{cases}
\label{cas11}
\eeq
where $C$ is a scalar in 2+1 dimensions and is a one-form
in 3+1 dimensions.
It is easy to see that setting the variation of $S_b$ with respect to 
$C$ to zero gives the vanishing of $F$ on the plates.
Thus in 3+1 dimensions, with the plate normal to the 3rd direction, for example,
we find $F_{14}= F_{24} = F_{12} = 0$, i.e., the vanishing of the tangential components of the electric field and the normal component of the magnetic field. These are the chromoelectric boundary conditions. We will comment on other boundary conditions later.

Notice that this action is invariant under the gauge transformations
$A \rightarrow g\, A\, g^{-1} - dg\, g^{-1}$,
$C \rightarrow g\, C\, g^{-1}$.
Further, in 3+1 dimensions, it is invariant under
$C \rightarrow C +  (d + A ) f$, by virtue of the Bianchi identity
on $F$. This is a new gauge symmetry for
$C$ and it shows that
we have only two independent components for
$C$, since one component can be removed (or related to the other
two) by a suitable choice
of $f$.
This would suggest that
we should expect only two independent boundary conditions
on $F$ upon carrying out the variation with respect to $C$.
This is indeed correct and is in complete agreement with the fact that
the Bianchi identity connects $F_{14},\, F_{24} $ and $ F_{12} $.
This also implies that the integration over $C$
 has to be done using a measure for $C$
which is invariant under this new gauge symmetry 
$C_\mu \rightarrow C_\mu + (\del + A)_\mu f$, i.e., a volume element
for the space of $C$'s modulo this gauge transformation.

One of the important consequences of the interactions
in the Yang-Mills theory will be
the dynamical generation of a propagator mass for the gluon.
As mentioned in the Introduction,
we expect that, for the range of $R$ we consider,
 a suitable approximation for the Casimir energy
 will be to neglect all further interactions once we take account of
 the dynamically generated mass.
 In other words, we expect that we can use a massive
 free propagator for the gauge field and also take
$F^a \approx d A^a$ in the boundary action
(\ref{cas11}). With this simplification, 
integrating out the $A$'s, we get $Z = {\cal N}\int [DC]\, e^{- \S_{\rm eff}}$
with
\beqar
\S_{\rm eff} &=& {1\over 2} \int (\del C_1)^{a\mu} \la A^a_\mu (\vx, 0) A^b_\nu (\vx',0)\ra (\del C_1 )^{b\nu}\nonumber\\
&&
+ {1\over 2} \int (\del C_2)^{a\mu} \la A^a_\mu (\vx, R) A^b_\nu (\vx',R)\ra (\del C_2 )^{b\nu}\nonumber\\
&&+ \int (\del C_1)^{a\mu} \la A^a_\mu (\vx, 0) A^b_\nu (\vx',R)\ra (\del C_2 )^{b\nu}
\label{cas12}
\eeqar
where
\beqar
&&(\del C_1)^{a\mu} = \epsilon^{\mu \alpha\beta} (\del_\alpha C^a_\beta
- \del_\beta C^a_\alpha )(\vx, 0)\nonumber\\
&&(\del C_2)^{a\mu} = \epsilon^{\mu \alpha\beta} (\del_\alpha C^a_\beta
- \del_\beta C^a_\alpha )(\vx, R)
\label{cas13}
\eeqar
Thus we need the propagators for $A$, with zero separation in
$x_3$ and with a separation of $R$. Because of the gauge symmetry of
$C$, we only need the gauge-invariant part of the propagators for
$A$. 

In 2+1 dimensions, we get the same action as in (\ref{cas12})
but $C$ is a scalar, so (\ref{cas13}) become
\beq
(\del C_1)^{a\mu} = \epsilon^{\mu \alpha} 
\del_\alpha C^a (\vx, 0),\hskip .2in
(\del C_2)^{a\mu} = \epsilon^{\mu \alpha} \del_\alpha C^a(\vx, R)
\label{cas14}
\eeq
with all indices $\alpha$, $\mu$ taking only two values corresponding to the 
tangential directions on the plates.

\section{Casimir energy in 2+1 dimensions}

We now turn to the details of the calculation of the Casimir energy,
starting with the (2+1)-dimensional case.
This is basically a review of our previous analytical calculation
of the Casimir energy in three dimensions \cite{KN1}
using the gauge-invariant Hamiltonian methods.

In this Hamiltonian analysis, we start with the gauge $A_0 =0$
and the spatial components of the gauge potentials 
were parametrized as \cite{{KKN1},{Ham2+1}}
\beq
A_z = {\half} (A_1 + i A_2) = - \del M ~M^{-1}, \hskip .2in A_{\bz} = {\half} (A_1 - i A_2 )
= M^{\dagger -1} \bdel M^\dagger
\label{1}
\eeq
Here we use complex coordinates $z = x_1 - i x_2$, $\bz = x_1 + i x_2$ with
$\del= {\half}( \del_1 + i \del_2)$, $\bdel = {\half} (\del_1 - i \del_2)$, and 
$M$ is an element of the complexified group $G^{\mathbb C}$.
Thus $M$ is an $SL(N, {\mathbb{C}})$-matrix
if the gauge transformations take values in $SU(N)$.
Gauge transformations act on $M$ via
$M \rightarrow M^g = g \,M$, where $g$ is an element of the group
$G$, say,  for example, $SU(N)$.

Wave functions are functionals of the gauge-invariant
combination $H = M^\dagger M$.
The inner product for the wave functions is
given as
\beq
\la 1\vert 2\ra = \int d\mu (H) \exp [2 c_{\rm A}~S_{\rm wzw}(H)]~ \Psi_1^* \Psi_2
\label{2}
\eeq
Here $S_{\rm wzw}$ is the Wess-Zumino-Witten action given by
\beq
S_{\rm wzw} (H) = {1 \over {2 \pi}} \int \Tr (\partial H ~\bdel
H^{-1}) +{i \over {12 \pi}} \int \epsilon ^{\mu \nu \alpha} \Tr (
H^{-1}
\partial _{\mu} H~ H^{-1}
\partial _{\nu}H ~H^{-1} \partial _{\alpha}H)
\label{3}
\eeq
Also $c_{\rm A}$ is the value of the quadratic Casimir operator
for the adjoint representation;
it is equal to $N$ for $SU(N)$.
Further, $d\mu (H)$  in (\ref{2}) is the Haar measure for 
$H$ which takes values in $SL(N, {\mathbb{C}})/SU(N)$.

The Hamiltonian and other observables can be expressed
in terms of the current $J$ of the WZW action, namely,
$J = {2 \over e} \, \del H ~ H^{-1}$.
It has the form
${\cal H} = {\cal H}_0 + {\cal H}_1$, with
\beqar
{\cal H}_0 &=& m  \int_z J_a (\vz) {\d \over {\d J_a (\vz)}} + {2\over \pi} \int _{z,w} 
 {1\over (z-w)^2} {\d \over {\d J_a (\vw)}} {\d \over {\d
J_a (\vz)}}\nonumber\\
&&\hskip .4in + {1\over 2} \int_x :\bdel J^a(x) ~\bdel J^a(x):\label{5}\\
{\cal H}_1& =&  i ~{e}~ f_{abc} \int_{z,w}  {J^c(\vw) \over \pi (z-w)}~ {\d \over {\d J_a (\vw)}} {\d \over {\d
J_b (\vz)}} \nonumber
\eeqar
where $m = e^2 c_{\rm A} /2\pi$. Regularization issues have been discussed
in the cited references.

Standard perturbation theory (in the Hamiltonian formulation), 
corresponds to expanding $H= \exp ( t_a \vf^a )$
in powers of the hermitian field $\vf^a$. Further, we should also expand
in powers of
$m = e^2 c_{\rm A}/2\pi$.
In our approach, we developed an expansion scheme
where we solve
the Schr\"odinger equation keeping all terms in ${\cal H}_0$
at the lowest  order, treating ${\cal H}_1$ as a perturbation. 
We also keep the term involving $m$
even at the lowest order.
Formally, we keep $m$ and $e$ as independent parameters in keeping track of different orders
in solving the Schr\"odinger equation, only setting $m = e^2 c_{\rm A}/2\pi$ at the end.
We may note that
 if we further expand $H$ in terms of $\vf^a$,
 our scheme would correspond to a partially resummed version 
 of  the usual perturbation expansion.

The lowest order result for the vacuum wave function was given in \cite{KKN1} and gave the string tension for a Wilson loop in the representation
${\rm R}$ as $\sigma_{\rm R} = e^4 c_{\rm A} c_{\rm R} /4 \pi$, $c_{\rm R}$ being the quadratic Casimir value
for the representation ${\rm R}$.
Corrections to this formula were considered in \cite{KNY}, taking the expansion to the next higher order (which still involves an infinity of correction terms) and found that the corrections were small, of the order of 
$-0.03\%$ to $-2.8\%$. The resulting values for the string tension
agree well with the lattice estimates
\cite{{teper},{teper2}}.\footnote{For considerations on glueball masses in this
approach, see \cite{LMY}.}

For the present purpose of obtaining the two-point function for
$A_\mu$,  it is useful to write the Hamiltonian in a different form
\cite{KNrobust}.
We can absorb the factor $e^{2 c_{\rm A} S_{\rm wzw}}$ in (\ref{2}) into the definition of the wave function by writing $\Psi = e^{- c_{\rm A} S_{wzw}} \, \Phi$.  
However, the
Hamiltonian acting on $\Phi$ will now be given by
${\cal H} \rightarrow e^{ c_{\rm A} S_{\rm wzw}}\, {\cal H}\, e^{- c_{\rm A} S_{\rm wzw}}$.
We then expand $H$ as  $H = \exp( t_a \vf^a ) \, \approx  1 + t_a \vf^a + \cdots$.
 This ``small $\vf$" expansion is suitable for the
intermediate kinematic regime of interest, it corresponds to
a (resummed) perturbation theory. The Hamiltonian is then
\beq
{\cal H}= {1\over 2}\int \left[ -{\delta^2 \over \delta \phi^2} +\phi (-\nabla^2 +m^2)\phi
+\cdots\right] \label{6}
\eeq
where $\phi _a (\vk) = \sqrt {{c_{\rm A} k \bar{k} }/ (2 \pi m)}~ \vf _a (\vk)$. 
This is clearly the Hamiltonian for a field of mass $m$ with the corresponding vacuum wave function 
\beq
\Phi_0 \approx \exp \left[ - {1\over 2} \int \phi^a \sqrt{ m^2 - \nabla^2} ~\phi^a \right]
\label{7}
\eeq
(Since we absorbed the factor $e^{2 c_{\rm A}~S_{\rm wzw}(H)}$
into the wave functions,
the inner product for the $\Phi$'s will involve
just the Haar measure, as is clear from (\ref{2}).)
We can view $\H$ in (\ref{6}) as arising from the  action
\beq
S = \int d^3x \, {1\over 2}\left[  {\dot \phi}^a {\dot \phi}^a - {(\nabla \phi^a ) (\nabla \phi^a ) }
- {m^2 \phi^a \phi^a }  \right] + \cdots
\label{8}
\eeq
This shows that, effectively, we have ${\rm dim}\,G$ 
copies of a massive scalar 
field, the mass being $m = e^2 c_{\rm A}/(2\pi)$.

Notice that for small $\vf$, for the gauge-invariant part of the
potential we have
$A^a_i \approx -{i \over 2}\e_{ij} \del_j \vf^a$.
What we have shown is that
$\vf^a$, modulo the multiplicative factor
$\sqrt {{c_{\rm A} k \bar{k} }/ (2 \pi m)}$,
 is described by a massive scalar field.
 It is possible to use this to construct the two-point function
 for $A_\mu$ and use it in (\ref{cas12}) to obtain the boundary
 action and obtain the Casimir energy by integrating over the
 $C$'s. However, in this case, the problem is simpler, since
 the boundary condition of vanishing tangential electric field is
 \beq
\epsilon_{ij} \, n_i F_{0j}^a = 0 ,
\label{8a}
\eeq
where $n_i$ is the unit vector normal to the
plate.
In terms of $\vf^a$, this becomes
\beq
n _i \epsilon_{ij} \epsilon_{jk} \del_k {\dot \vf^a} = 
- n_i \del_i {\dot \vf^a} = 0
\label{8b}
\eeq 
This
is equivalent to imposing Neumann boundary conditions on the
scalar field $\vf^a$. The field $\phi^a$ will obey the same condition.
We can then, in a straightforward way, expand $\phi^a$ in terms of modes
obeying the Neumann condition and calculate the Casimir energy.
This was carried out in \cite{KN1}, we will therefore just quote the
result.
The Casimir energy per unit length
is given by
\beq
{\E \over L} = - {{\rm dim}G \over 16\pi R^2}  
\left[  2m R ~ {\rm Li}_2 (e^{- 2 m R}) + {\rm Li}_3(e^{- 2 m R} )\right]
\label{12a}
\eeq
where ${\rm Li}_s (w)$ is the polylogarithm function
\beq
{\rm Li}_s (w) = \sum_{1}^\infty  {w^n \over n^s}
\label{12}
\eeq
For ease of comparison with the lattice-based numerical calculations,
it is useful to express this in terms of the string tension 
$\sigma_{\rm F}$ for the fundamental
representation. For $SU(N)$, we have
\beqar
\sigma_{\rm F} &=& e^4 {c_{\rm A} c_{\rm F} \over 4 \pi}
= e^4 \, {N^2 -1 \over 8 \pi}\nonumber\\
m &=& {e^2 c_{\rm A}\over 2 \pi} = \sqrt{c_{\rm A} \over \pi c_{\rm F}}\,\sqrt{\sigma_{\rm F}} = \sqrt{2 N^2 \over \pi (N^2-1)} \, \sqrt{\sigma_{\rm F}}
\label{12b}
\eeqar
where we used $c_{\rm A} = N$, $c_{\rm F} = (N^2 -1) / (2 N)$.
In terms of $\sigma_{\rm F}$,
the Casimir energy per unit length is thus
\beq
{\E \over L \sigma_F } = - {{\rm dim}G \over 16\pi} \left[ {2\sqrt{c_{\rm A}/\pi c_{\rm F}} \over x}\,
{\rm Li}_2 \left( e^{- 2\sqrt{c_{\rm A}/\pi c_{\rm F}} \,x} \right) + { 1 \over x^2} {\rm Li}_3 \left( e^{- 2\sqrt{c_{\rm A}/\pi c_{\rm F}}\, x} \right)
\right]
\label{13}
\eeq
where $x = R\sqrt{\sigma_{\rm F}} $.

For very small distances, $R \ll m^{-1}$, the Casimir energy (\ref{13}) is that of 
${\rm dim}G$ massless scalar fields \cite{KN1},
\beq
{\E \over L \sigma_F } \rightarrow - {\rm dim}G\,{\zeta(3) \over 16 \pi x^2}
\label{13a}
\eeq

The boundary condition (\ref{8a}) or (\ref{8b}) corresponds to
the case where both wires are perfect electric conductors
(PEC). In \cite{KN1}, the Casimir energy was calculated for
other boundary conditions as well. For example, the case where the wires are 
perfect magnetic conductors (PMC), i.e., with the magnetic field vanishing on
the wires, is realized by the field $\phi^a$ obeying Dirichlet conditions. 
The Casimir energy for scalar fields obeying Dirichlet or Neumann conditions on both plates is the same (\ref{12a}), so we can conclude that the Casimir energy for 
PMC conditions on both wires
is the same as for PEC, i.e., $\E_{\rm PMC} = \E_{\rm PEC}$.
For the case of one wire with the PEC condition and the other with
the PMC condition, the Casimir energy is positive, corresponding to
a repulsive Casimir force \cite{KN1}.
More general boundary conditions as in \cite{{KN2},{asorey}}
can also be considered. However, since
there is no available data on these from lattice simulations, we do not quote these results here. But it is worth pointing out that
lattice simulations with boundary conditions other than the perfect electrical conductor will be very interesting, providing additional points of comparison.

 \section{Casimir energy in 3+1 dimensions}
 
 The key ingredient which facilitated the Hamiltonian analysis in
 2+1 dimensions was the parametrization of the gauge potentials
 as in (\ref{1}). A simple argument for this uses the fact that
 we can think of the spatial manifold
 as the large radius limit of a two-sphere $S^2 = SU(2)/U(1)$. 
Since the representation matrices or Wigner $\D$-functions
of $SU(2)$ constitute a complete
set, we can then obtain the parametrization (\ref{1}) \cite{{Ham2+1},{AN}}.
The complex structure of $S^2$ is also crucial for this result.
The manifold $\mathbb{CP}^2 = SU(3) /U(2)$ is a suitable
choice for a similar strategy in four dimensions.
Since this has the required number of spacetime dimensions,
the appropriate approach would be to use
a Euclidean functional integral, rather than a Hamiltonian analysis.
The parametrization of the gauge fields is of the form
\cite{KMN}
\beqar
A_i &=& -\nabla_i M M^{-1} - M a_i M^{-1}, \hskip .2in
\bA_{\bar i} = M^{\dagger -1} \bnabla_{\bar i} M^\dagger 
+ M^{\dagger -1} \ba_{\bar i} M^\dagger
\label{4d-1}\\
a_i &=&
- g_{i {\bar i}} \e^{{\bar i} {\bar j}} \left({\bnabla}_{\bar j} \chi
+ [ H^{-1} \nabla_{\bar j} H,  \chi]\right)\nonumber\\
\ba_{\bar i} &=&   - g_{{\bar i} i } \e^{ij} 
\left( \nabla_j \chi^\dagger + [ - \nabla_j H H^{-1} , \chi^\dagger ]
\right)  = {a_i}^\dagger
\label{4d-2}
\eeqar
$M$ is again a complex matrix taking values in the complexification of the
group $G$. $\chi$ is a complex matrix field, but it is
not a scalar, rather $\chi \e^{{\bar i} {\bar j}}$ is a rank-2 antisymmetric
tensor under $U(2)$. (While there is only a single complex field 
$\chi$, it is not a scalar, it transforms nontrivially under 
the `hypercharge' $U(1)\in U(2) \sim SU(2) \times U(1)$, see \cite{KMN}.)
The gauge-invariant variables are
$H = M^\dagger M$ and $\chi$ and $\chi^\dagger$.

Since gauge transformations act as before, namely,
$M \rightarrow M^g = g \, M$, it is possible to factor out the gauge degrees of
freedom and define a functional measure of integration
on the gauge-orbit space.
However, unlike the lower dimensional case, the (functional) Jacobians
for the change of variables $A_i, \, {\bA_{\bar i}} \rightarrow H, \chi, \chi^\dagger$ cannot be calculated exactly.
We have calculated the leading terms in a derivative expansion
of the fields. The key result is that
there is a WZW action with a finite coefficient and
the fields $a_i$ and ${\bar a}_{\bar i}$
acquire mass \cite{KMN}.
However, unlike the lower dimensional case, this mass is an input parameter.
Recall that to define the quantum Yang-Mills theory in four dimensions
we need an input parameter with the dimension of mass.
In perturbative calculations, this is usually taken as $\Lambda_{\rm QCD}$.
In our nonperturbative set-up, the dimensional transmutation is
obtained via the mass parameter for $a_i$, ${\bar a}_{\bar i}$.
If we were to calculate the one-loop corrections, then
we could obtain $\Lambda_{\rm QCD}$ as a function of this mass
parameter but we have not completed this yet.

We can simplify the parametrization (\ref{4d-1})
in the large radius limit of $\mathbb{CP}^2$.
For our purpose it is sufficient to approximate $M$ as $M \approx 1 +i \theta - \vf_3$. The hermitian and Cartesian components of the fields then take the form
\beq
A^a_4 = \del_4 \theta^a + \nabla\cdot \vf^a,
\hskip .2in
A_i^a = \del_i \theta^a - \del_4 \vf_i^a - \e_{ijk} \del_j \vf_k^a
\label{4d-3}
\eeq
Evidently, up to linear order of gauge transformations, the gauge degrees of freedom are $\theta^a$; 
in addition, we have
three sets of fields $\vf_i^a$, $i =1, 2, 3$, with
$\chi$ in (\ref{4d-2}) given by
$\chi = - \vf_1 - i \vf_2$.

In order to use our theory for Casimir energy calculations in $\mathbb{R}^4$, we consider the large radius limit of $\mathbb{CP}^2$, effectively reproducing flat space. 
However, recall that the large radius limit of 
$\mathbb{CP}^2$ is not $S^4$ but a compactified version of
$\mathbb{C}^2$, so that we do not obtain full rotational symmetry. 
We can covariantize our results in complex space by integrating over the placements
of $\mathbb{C}^2$ in $\mathbb{R}^4$. This recovers the full rotational symmetries, leading to results relevant to standard $\mathbb{R}^4$. 
(This is similar to what is done with twistors where one eventually integrates
over all local complex structures on $S^4$.)
The fields $\vf^a_i$ acquire dynamically generated masses as explained above. Under covariantization, all $\vf_i^a$ have the same mass (which depends on the mass parameter for $a_i, \ba_{\bi}$ and the finite coefficient of the WZW action). The action for the $\vf$-fields becomes
\beq
\S = {1\over 2} \int \left[ (\Box \vf^a_i)^2 + m^2 (\del \vf^a_i)^2
\right]
\label{4d-4}
\eeq
For more details leading to this result, see Appendix.

In terms of gauge fields, up to quadratic order in $\vf^4_i$, the action (\ref{4d-4}) can be written as 
\beq
S = \int \left[{1\over 4} F_{\mu\nu} F^{\mu\nu} + {m^2 \over 4}
F_{\mu\nu} {1\over (-\Box)}\, F^{\mu\nu} +\cdots\right]
 \label{4d-4a}
\eeq

As mentioned in section 2, we need the gauge-invariant part of the two-point
function for $A$'s. Given the parametrization of the fields as in
(\ref{4d-3}), it is convenient in this case to take the plates to be orthogonal
to the $x_4$-direction, so that we have $A_i$, $i =1,2,3$, in the boundary action. (The mapping $(1234) \rightarrow (4123)$ will connect these results
back to the case in section 2.)
Further, $\theta$ is irrelevant for our purpose, so we need
\beq
\la A_i^a (x) A_j^b (y) \ra =
\big\langle (\del_4 \vf_i^a + \e_{ipq} \del_p \vf_q^a)(x)\,
(\del_4 \vf_j^b + \e_{jrs} \del_r \vf_s^b)(y)\big\rangle
\label{4d-5}
\eeq
For this correlation function, the basic result we need from the action
(\ref{4d-4}) is
\beq
\la \vf^a_i (x)\, \vf^b_j (y) \ra = \delta^{ab} \delta_{ij} \, G(x, y) 
= \delta^{ab} \delta_{ij} \int {d^4 k \over (2\pi)^4} {e^{-i k(x-y)} \over k^2 (k^2
+ m^2)}
\label{4d-6}
\eeq
Using this result it is easy to see that
\beqar
\la A_i^a (x) A_j^b (y) \ra &=& \delta^{ab}
\left[ -\del_4^2 \delta_{ij} - \e_{ipj} \del_p \del_4 -
\e_{jpi} \del_p \del_4 - ( \delta_{ij} \delta_{pr} - \delta_{ip} \delta_{jr} ) \del_p \del_r \right] G(x,y)\nonumber\\
&=& \delta^{ab} \left[ - \Box \delta_{ij} + \del_i \del_j \right] G(x,y)
\label{4d-7}
\eeqar
This can now be used in (\ref{cas12}) to obtain the effective boundary action
as
\beqar
\S_{\rm eff} &=&{1\over 2} \int (C_1^T)^{ai}(x) \, K_{00} (x, y)\,(C_1^T)^{ai}(y) 
+{1\over 2} \int (C_2^T)^{ai}(x) \, K_{\rm RR} (x, y)\,(C_2^T)^{ai}(y) \nonumber\\
&&+ \int (C_1^T)^{ai}(x) \, K_{\rm 0R} (x, y)\,(C_2^T)^{ai}(y) 
\label{4d-8}
\eeqar
In carrying out this simplification, we note that we can separate
$C^{ai}$ into longitudinal (proportional to $\del^i$ of some function)
and transverse components. The longitudinal component does not
appear in $\S_{\rm eff}$ because (\ref{4d-7}) shows that the 
correlation function is transverse. 
For the transverse components, the $\del_i \del_j$ part in
(\ref{4d-7}) gives zero. An integration by parts transfers the derivatives to the correlator from the $C$'s. This gives the expression (\ref{4d-8}).
We have used the superscript $T$ on $C$ to emphasize that it is only
the transverse component which appears here. Further the two-point
correlation functions
in (\ref{4d-8}), after $k_4$-integration, are given by
\beqar
K_{00} (x, y) &=& \left[ \nabla^2 \Box G(x,y) \right]_{x_4 = y_4 = 0}
= \int {d^3k \over (2\pi)^3} {\vk^2 \over 2 \omega_k } \, e^{- i \vk \cdot (\vx-\vy)}\nonumber\\
K_{\rm RR} (x, y) &=& \left[ \nabla^2 \Box G(x,y) \right]_{x_4 = y_4 = R}
= K_{00}(x, y)
\label{4d-9}\\
K_{\rm 0R} (x, y) &=& \left[ \nabla^2 \Box G(x,y) \right]_{x_4 =0, y_4 = R}
= \int {d^3k \over (2\pi)^3} {\vk^2 e^{- \omega_k R} \over 2 \omega_k } \, 
e^{- i \vk \cdot (\vx-\vy)}
\nonumber
\eeqar
where $\omega_k = \sqrt{\vk^2 + m^2}$.

The integration over $(C_1^T)^{ai}$ is now straightforward.
Because of the transversality condition, there are only
two independent fields for each of $C_1^T$ and $C_2^T$.
To be explicit, one could even decompose these fields as
\beq
(C_1^T)^{ai} = \int d^3x \left[ i \psi^a_1(q)  \, e_{(1)}^ i (q) + \psi_2 (q) \, e_{(2)}^i (q) \right] \, e^{- i \vq\cdot \vx}
\label{4d-10}
\eeq
where $\psi_1$, $\psi_2$ are independent fields and
$e_{(1)}^ i (q)$ and $e_{(2)}^ i (q)$ are two polarization vectors transverse to
$\vec q$. One can write a similar expansion for $(C_2^T)^{ai}$ as well.
Integrating over the $C$'s we find the $R$-dependent terms in $\log Z$ to be 
given by
\beq
-\log Z = {\rm dim}G~ \Tr \log  \Bigl(1 - K^{-1}_{\rm RR} K_{\rm R0} K^{-1}_{00} K_{\rm 0R} \Bigr)
\label{4d-11}
\eeq
There is no factor of $\half$ as in (\ref{cas8}) since there are
two polarizations which contribute.
Using (\ref{4d-9}) we find
\beqar
- \log Z &=&{\rm dim}G\,  \int d^3x {d^3k \over (2\pi)^3} \log \left( 1- 
e^{- \omega_k R}\right)\nonumber\\
&=&- {\rm dim}G\,  \int d^3x {d^3k \over (2\pi)^3} \sum_n {e^{-n \omega_k R} \over n}
\nonumber\\
&=& -{\rm dim}G\,  \int d^3x \sum_n {m^3 \over 8 \pi^2}\int_0^\infty ds \, (\cosh 3 s - \cosh s )
{e^{ - 2 m R n \cosh s}\over n}
\label{4d-12}
\eeqar
where we have used the substitution $\vert \vk \vert = m \sinh s$
and carried out the angular integrations.
We can now use the integral representation of the modified Bessel function
\beq
K_\nu (z) = \int_0^\infty ds\, e^{- z \cosh s} \, \cosh (\nu s)
\label{4d-13}
\eeq
to simplify (\ref{4d-12}) as
\beqar
- \log Z &=& - {\rm dim}G\,   \int d^3x \,\sum_n {m^3 \over 8 \pi^2 n} \Bigl( K_3 (2 m R n)
- K_1 (2 m R n) \Bigr)\nonumber\\
&=&-{\rm dim}G\,  \int d^3x \,{m^2 \over 4 \pi^2 R} \sum_n  {K_2 (2 m R n) \over n^2}
\label{4d-14}
\eeqar
where we have also used the identity
\beq
K_3 (z) - K_1 (z) = {4 \over z} \, K_2 (z)
\label{4d-15}
\eeq
The Casimir energy per unit area of the plates is thus given by
\beq
{\E \over \A (\sigma_{\rm F})^{3\over 2}}
= - C\,{\rm dim}G \,{(\tilde{m})^2\over 4 \pi^2\, x}
\sum_n {K_2 ( 2 \tilde{m} x n )\over n^2}
\label{4d-16}
\eeq
where $x = R \sqrt{\sigma_{\rm F}}$ and 
${\tilde m} = m /\sqrt{\sigma_{\rm F}}$,
with $\sigma_{\rm F}$ being the string tension for the fundamental representation. We have also included a prefactor $C$. This is actually $1$ for us, but we include it to facilitate comparisons and comments later.

The final result is equivalent to what we would find from the exchange of
$2\, {\rm dim}G$ massive scalar fields between the plates. 
(For the calculation of the energy for massive scalars, see also
\cite{{casimir},{mass-scalar}}.)

In the limit of small $R m = {\tilde m} x$ (i.e., at short distances 
$x \ll {\tilde m}^{-1}$ or in the limit of $m\rightarrow 0$),
the modified Bessel function can be approximated by
\beq
K_{2} ( 2 {\tilde m} x n) \sim {1\over 2 {\tilde m}^2 n^2 x^2} + \cdots
\label{4d-16a}
\eeq
Using this, the energy formula (\ref{4d-16}) becomes
\beq
{\E \over \A (\sigma_{\rm F})^{3\over 2}}
= - C\,{\rm dim}G \, {1\over 8 \pi^2 x^3 } \sum_n {1\over n^4}
= - C\,{\rm dim}G \, {\pi^2 \over 720 x^3}
\label{4d-16b}
\eeq
This agrees (with $C =1$ ) with result for $ 2\, {\rm dim}G$
massless scalar fields. (We have normalized the result
such that $C=1$ corresponds to two polarizations.)

The result (\ref{4d-16}) is also what was used
as an ansatz in \cite{chern2} in analyzing their lattice data.
However, we emphasize that, for us, it is not an ansatz,
 it follows from our gauge-invariant
analysis for the 4d YM theory.

It is also interesting to consider other types of boundary conditions
on the fields. The conditions we used, i.e., the vanishing of the
components of $F^a_{\mu\nu}$ tangential to the plates, correspond to a
perfect (chromo)electric conductor (PEC).
The vanishing of the dual components, namely,
$F^a_{14} = F^a_{24} = F^a_{34} = 0$ (or $F^a_{34}= F^a_{13} = F^a_{23}
= 0$ in the choice of axes used in section 2)
is another possibility, corresponding to the perfect (chromo)magnetic
conductor (PMC). For this case, one might consider the boundary action
\beq
S_b = i \int C^a_i \, F^a_{i 4}
\label{4d-16c}
\eeq
In the PEC case, the Bianchi identity among the components of
$F$ in the boundary term in (\ref{cas11}) led to a new gauge symmetry
$C \rightarrow C + \del f$, thus effectively reducing $C$ to two
polarizations. If we use (\ref{4d-16c}) though, all three polarizations of $C$
will in principle contribute. The Casimir energy will thus be $3\over 2$ times the result for the PEC conditions \cite{saharian, dudal}.  A similar derivation in 2+1 dimensions would imply that the Casimir energy for PMC conditions would be 2 times the result for the PEC conditions, which would be contradictory to the results we found in section 3. The Hamiltonian analysis in section 3 showed that the Casimir energy for both PEC and PMC conditions is the same. 

We believe that the resolution of this issue has to do with the nature of mass $m$. 
The mass used here is dynamically generated and therefore
we do not expect it to be a hard mass, although one can use the action
(\ref{4d-4}) for a certain kinematic regime.
At high energies or short distances, we should also expect
the theory to reduce to a theory of massless gluons, 
by virtue of asymptotic freedom.
But a hard mass, however small,
should have three polarizations just by representation theory for
the Poincar\'e symmetry. Therefore, with a hard mass,
generally we will not obtain a smooth limit
to the zero mass case at short distances.
(The PEC case does correspond to two polarizations in 
the short distance limit,
as is clear from (\ref{4d-16b}).)
The dynamical mass on the other hand will go to zero at high energies,
so it is possible to get a result that agrees with two contributing
polarizations. The likely scenario is that even if we use the hard mass
as in (\ref{4d-4}) for the kinematic regime relevant to the Casimir
energy simulations, we should have two polarizations
contributing.

The result for the Casimir energy, for massless electrodynamics  is the same for both PMC and PEC boundary conditions and results from the 
contribution of
two polarizations, see \cite{MIT-piston, dudal1}. This indicates that
(\ref{4d-16c}) is not the right boundary action to be used for the PMC conditions in the case of dynamical mass, and has to be modified for consistency with the short distance limit. One way to do that is to notice that if we have a mode expansion
of the fields where $F_{i4} = 0$, by the Bianchi identity, these
give $\del_4 F_{12} = \del_4 F_{23} = \del_4 F_{31} = 0$.
By the same Bianchi identities, these are equivalent to
$\epsilon^{ijk} \del_j F_{4k} = 0$. Therefore, 
an alternate, perhaps better, possibility is to use
the boundary action
\beq
S_b = i \int C^a_i \epsilon^{ijk} \del_j F^a_{4k}
=  i \int C^a_i \epsilon^{ijk} \del_j F^a_{\mu k}\, n^\mu
\label{4d-16d}
\eeq
as the action relevant to the PMC conditions.\footnote{Another choice would be
\beq
S_b = i \int C^a_i \epsilon^{ijk} \del_4 F^a_{jk}
=  i \int C^a_i \epsilon^{ijk} \del_\mu F^a_{j k}\, n^\mu .
\label{4d-16dd}
\eeq }
 (Here $n^\mu$ is the unit vector
normal to the plate.)
In this case, we see that we again have the gauge symmetry
$C \rightarrow C+ \del f$; thus $C$ has only two polarizations and the
result of calculating with (\ref{4d-16d}) gives the same result
for the Casimir energy as the PEC conditions. An equivalent way would be to use the original boundary action (\ref{4d-16c}) but impose by hand a transversality condition for $C$. 

Again, lattice simulations with PMC boundary conditions would be very useful in establishing and clarifying the relation for the Casimir energy for PEC and PMC boundary conditions in the truly nonperturbative setting of Yang-Mills theory.

\section{Comparison with numerical calculations}

We will now compare our analytical formulae 
(valid for any gauge group)
with the results for the numerical calculation via lattice simulations of the Casimir energy for the nonabelian gauge theory. The lattice simulations were performed for perfect chromoelectric boundary conditions (PEC case).
Results are available for the groups $SU(2)$ and $SU(3)$ in both
three (i.e., 2+1) and four (i.e., 3+1) dimensions. We will consider each dimension separately.

\subsection{Three dimensions}

The first lattice calculation for the Casimir energy for $SU(2)$ was 
given in \cite{chern1}.
The authors fit their results to the formula
\beq
{\E \over L \sigma_{\rm F}} = - {\rm dim}G\, { \zeta (3)\over 16 \pi}\, x^{-\nu} \,
e^{- M_{\rm Cas}\, x / \sqrt{\sigma_{\rm F}}}
\label{com1}
\eeq
The best-fit values of the parameters were
$\nu = 2.05$ and $M_{\rm Cas} = 1.38\, \sqrt{\sigma_{\rm F}}$.
More recently, there has been another calculation for $SU(2)$
\cite{ngwenya}.
The authors again fit their results to a formula of the form
(\ref{com1}), with
 $\nu = 2.022$ and $M_{\rm Cas} = 1.38\, \sqrt{\sigma_{\rm F}}$.

In our view, and as we have already argued in \cite{KN1}, 
there is stronger motivation for our formula (\ref{13}) as it
follows from a Hamiltonian analysis based on first principles
which has also produced a prediction for the string tension
in close agreement with lattice-based measurements.
Further, in (\ref{13}), there is a {\it predicted} value for the mass parameter
as well, see (\ref{12b}).
We may also note that there is a difference in interpretation.
For us, the relevant degrees of freedom are the gluons with a
propagator mass, we are not considering new excitations.

\begin{figure}[!b]
\begin{center}
\begin{tikzpicture}[scale = 1.4,domain=0:7]
\pgftext{
\scalebox{.4}{\includegraphics{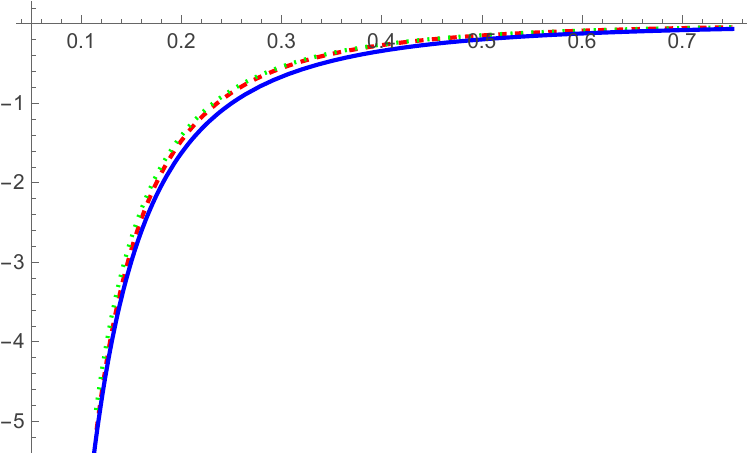}}
\scalebox{.4}{\includegraphics{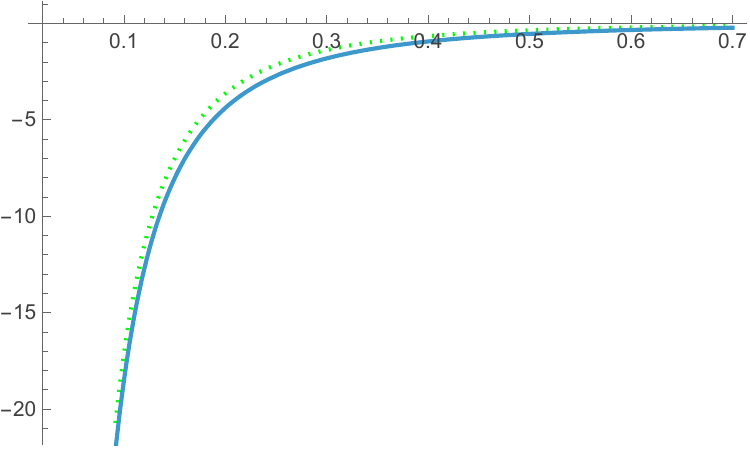}}
}
\draw(-5.4,.5)node{${\E\over L\sigma_{\rm F}}$};
\draw(-3.7,1.6)node{$x\,\rightarrow $};
\draw(1.8,1.6)node{$x\,\rightarrow $};
\end{tikzpicture}
\caption{Comparison of lattice-based results from \cite{chern1}
(red,dashed), from \cite{ngwenya}(green, dotted) with our
formula (\ref{13}) for $SU(2)$ on the left and $SU(3)$ on the right.}
\label{casgraph1}
\end{center}
\end{figure}
The comparison of these cases is given in
Fig.\,\ref{casgraph1}. The red dashed line shows the
formula (\ref{com1}) for 
$\nu = 2.05$ and $M_{\rm Cas} = 1.38\, \sqrt{\sigma_{\rm F}}$, corresponding to
\cite{chern1}, the green dotted line is for
$\nu = 2.022$ and $M_{\rm Cas} = 1.38\, \sqrt{\sigma_{\rm F}}$
corresponding to \cite{ngwenya}.
The blue solid line is our result (\ref{13}), with no adjustable 
parameters. We note that the mass which controls the exponential decay
in our formula (\ref{13}) is
$2m = 2 \sqrt{c_{\rm A} /\pi c_{\rm F}} \sqrt{\sigma_{\rm F}} \approx
1.84\, \sqrt{\sigma_{\rm F}}$ for $SU(2)$,
from (\ref{12b}).

The lattice-based estimate of the energy has been obtained
for the case $SU(3)$ in \cite{ngwenya}. The authors
 again fit the results to the formula
(\ref{com1}), the best-fit values being $\nu = 2.02$, $M_{\rm Cas} = 1.51\sqrt{\sigma_{\rm F}}$.
We show this graph in Fig.\,\ref{casgraph1} on the right side.
Our result (\ref{13}) is again shown in the blue solid line.
In this case, we note that $2 m = \sqrt{9/\pi} \sqrt{\sigma_{\rm F}}
\approx 1.69\, \sqrt{\sigma_{\rm F}}$.

In both cases, we note that the result from the
Hamiltonian analysis is very close to the lattice-based numerical
calculations, with no adjustable parameters.

Another feature about the formula (\ref{com1}) {\it vis-\`a-vis} 
our result (\ref{13}) which is worthy of comment is the following.
Notice that best-fit value of
$M_{\rm Cas}$ in units of $\sqrt{\sigma_{\rm F}}$
is higher for $SU(3)$ compared to $SU(2)$.
In the same units, the lowest lying glueball masses
are $4.716$ for $SU(2)$ and $4.330$ for $SU(3)$
\cite{LucT}.
It is a bit peculiar that while the glueball mass decreases
as we go from $SU(2)$ to $SU(3)$, $M_{\rm Cas}$ increases.
This would be hard to understand if any excitation corresponding to
$M_{\rm Cas}$ is part of the glueball.
On the other hand, for us, the value of $2 m$ decreases 
from $1.84$ to $1.69$ as we go from
$SU(2)$ to $SU(3)$ in consonance with 
the behavior of glueball masses.

\subsection{Four dimensions}

The situation in four dimensions is more involved.
The Casimir energies have been calculated by
 lattice simulations for $SU(2)$
in \cite{ngwenya} and for $SU(3)$ by \cite{chern2} and 
by \cite{ngwenya}.
For $SU(2)$, the authors of \cite{ngwenya} fit their results to a formula of the form
\beq
{\E \over \A (\sigma_{\rm F})^{3\over 2}} = - {\rm dim}G\, { \zeta (4)\over 16 \pi^2}\, x^{-\nu} \,
e^{- M_{\rm Cas}\, x / \sqrt{\sigma_{\rm F}}}
\label{com2}
\eeq
The best-fit choice of parameters is
$\nu = 3.09$ and $M_{\rm Cas} = 0.38 \sqrt{\sigma_{\rm F}}$ for $SU(2)$.

\begin{figure}[!b]
\begin{center}
\begin{tikzpicture}[scale = 1.4,domain=0:7]
\pgftext{
\scalebox{.5}{\includegraphics{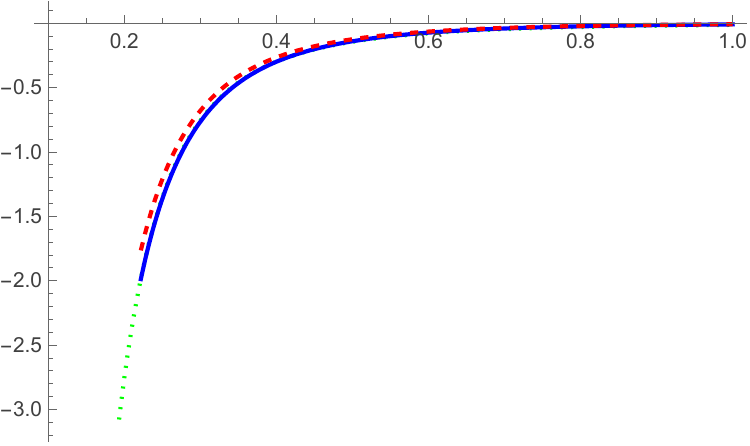}}
}
\draw(-3.7,.5)node{${\E\over \A(\sigma_{\rm F})^{3/2}}$};
\draw(0,1.9)node{$x\,\rightarrow $};
\end{tikzpicture}
\caption{Comparison of lattice-based results for $SU(2)$ in 4 dimensions
from \cite{ngwenya}
(green, dotted) with our
formula (\ref{4d-16}) for $SU(2)$, with $m = 1.067 \sqrt{\sigma_{\rm F}}$
 and $C = 1.126$ (red, dashed) and $C = 1$ (blue, solid).}
\label{casgraph2}
\end{center}
\end{figure}

We consider the values given by the function (\ref{com2}) for the range
$.2\leq x \leq 1$ and fit to our formula (\ref{4d-16}).
The best-fit values are $C = 1.126$ and $m = 1.067 \sqrt{\sigma_{\rm F}}$
(i.e., ${\tilde m} = 1.067$).
However, our calculation gave $C = 1$. So in Fig.\,\ref{casgraph2},
we show the curve corresponding to (\ref{com2}) with
$\nu = 3.09$, $M_{\rm Cas} = 0.38 \sqrt{\sigma_{\rm F}}$
(corresponding to lattice data)
and our formula (\ref{4d-16}) with $C= 1.126$ (red, dashed) , $C =1$ (blue, solid) and
$m = 1.067 \sqrt{\sigma_{\rm F}}$.
Notice that all three curves are essentially identical.

It is also possible to find a fit for a larger range,
$0.1\leq x \leq 1$, with somewhat different parameters.
But we have chosen the range $0.2\leq x\leq 1$
for the graph of our formula shown in 
Fig.\,\ref{casgraph2} since the lattice data
for \cite{chern2} is in this range.
We note however that,
as mentioned earlier, we expect the analysis to hold
only for a certain range, where the dynamical mass generation effects are
important, but the degrees of freedom are still gluons rather than
glueballs, but it is difficult to know {\it a priori} what this range should be.

As mentioned, for $SU(3)$ there are calculations from
\cite{chern2} and \cite{ngwenya}.
The authors of \cite{chern2} fit the results to a formula of the same form
as we have argued for, namely (\ref{4d-16}), but with a prefactor $C \neq 1$.
Although this is basically the formula we also have, we emphasize that
the merit of our gauge-invariant analysis is really to justify 
the use of a formula of this form.
The best-fit values for \cite{chern2} give
$C = 5.6 $ and $m = 1.0 \sqrt{\sigma_{\rm F}}$.
The authors of \cite{ngwenya} fit to the formula
(\ref{com2}), the best-fit values for $SU(3)$ being
$\nu = 3.002$ and $M_{\rm Cas} = 0.06 \sqrt{\sigma_{\rm F}}$.

\begin{figure}[!t]
\begin{center}
\begin{tikzpicture}[scale = 1.4,domain=0:7]
\pgftext{
\scalebox{.5}{\includegraphics{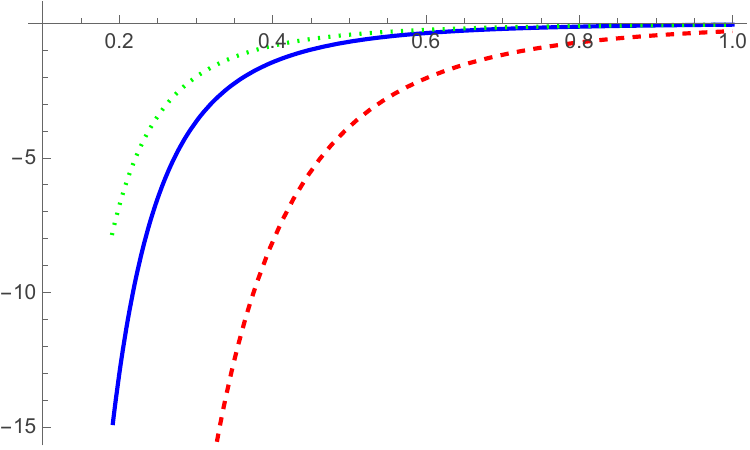}}
}
\draw(-3.7,.5)node{${\E\over \A(\sigma_{\rm F})^{3/2}}$};
\draw(0,1.9)node{$x\,\rightarrow $};
\end{tikzpicture}
\caption{Comparison of lattice-based results for $SU(3)$
in 4 dimensions from 
\cite{chern2} (red, dashed) and from \cite{ngwenya}
(green, dotted) with our
formula (\ref{4d-16}) for $SU(3)$, with $m = 1.0\sqrt{\sigma_{\rm F}}$
and $C = 1$ (blue, solid).}
\label{casgraph3}
\end{center}
\end{figure}

In Fig.\,\ref{casgraph3}, we show three curves, 
the red dashed curve for
 \cite{chern2} (from (\ref{4d-16}) with $C= 5.6, m =1$),
the green dotted curve for \cite{ngwenya} (from
(\ref{com2}) with $\nu = 3.002, M_{\rm Cas} = 0.06$) and
the blue solid curve for our formula (\ref{4d-16}) with
$C = 1$ and $m =1.0 \sqrt{\sigma_{\rm F}}$.

\begin{figure}[!t]
\begin{center}
\begin{tikzpicture}[scale = 1.4,domain=0:7]
\pgftext{
\scalebox{.5}{\includegraphics{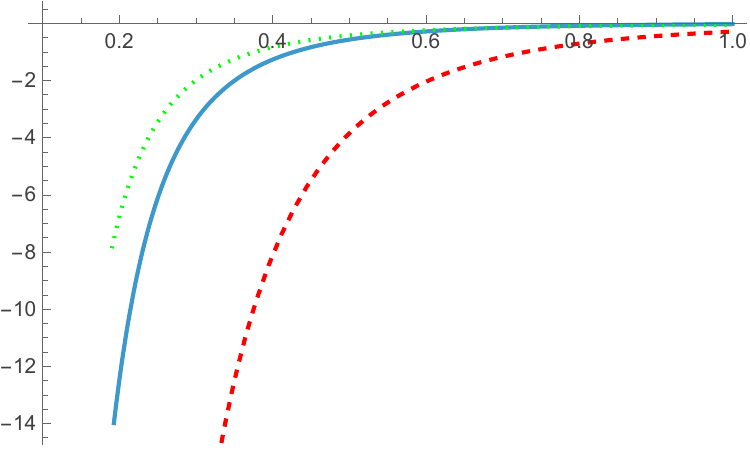}}
}
\draw(-3.7,.5)node{${\E\over \A(\sigma_{\rm F})^{3/2}}$};
\draw(0,1.9)node{$x\,\rightarrow $};
\end{tikzpicture}
\caption{Comparison of lattice-based results for $SU(3)$
in 4 dimensions from 
\cite{chern2} (red, dashed) and from \cite{ngwenya}
(green, dotted) with our
formula (\ref{4d-16}) for $SU(3)$, with $m = 1.41\sqrt{\sigma_{\rm F}}$
and $C = 1$ (blue, solid).}
\label{casgraph4}
\end{center}
\end{figure}
The key point here is that there is some disagreement 
between the two sets of lattice data.
If we use the mass value from \cite{chern2} and set $C= 1$, our
result
 is somewhere between the two lattice results, but given the 
discrepancy between the two lattice results, it is difficult to make
a definite conclusion.
Obviously if we allow for a
prefactor $C$ which is not necessarily $1$,
it is possible to fit our results to the data from \cite{chern2}, 
with $C = 5.6$ and $m = 1.0\sqrt{\sigma_{\rm F}}$, as is done in
\cite{chern2},
since the formulae are identical.
We can similarly fit our formula (\ref{4d-16}) to the lattice data from
\cite{ngwenya} for $SU(3)$ by choosing $C= 0.5$ and
$m = 0.38 \sqrt{\sigma_{\rm F}}$.

Another comparison one might consider is the following.
There is by now evidence, based on lattice calculations
and arguments based on Schwinger-Dyson equations, that the gluon propagator acquires a mass \cite{masslat}-\cite{papavassiliou}.
 In the latest analyses, for example, the
value obtained for $SU(3)$ is $m \approx 686\, {\rm MeV}
 \approx 1.41 \sqrt{\sigma_{\rm F}}$, where we use
 $\sqrt{\sigma_{\rm F}} \approx 0.485 \,{\rm GeV}$ \cite{papavassiliou}.
If we use this value and calculate the energy according to
our formula (\ref{4d-16}), with $C =1$, the result is again
between the two sets of lattice data, see Fig.\,\ref{casgraph4}.
Again, an unambiguous set of lattice data 
is needed before we can conclude anything about possible
compatibility with other methods
of calculation.

A comment on the prefactor $C$ is appropriate at this stage.
The energy is given by a determinant which arises
from a functional integration, so it is easy to see that $C$ is
related to the number of degrees of freedom which can contribute.
Our result (\ref{13}) corresponds to one field degree of freedom
for each direction in the Lie algebra (or one
polarization) and result (\ref{4d-16}) corresponds,
with $C = 1$, to two polarizations. This is in agreement with
general expectations for the gauge theory in 2+1 and 3+1 dimensions.
In principle, $C$ cannot therefore be a parameter
for which we seek a best-fit value.
Nevertheless, there is some variance possible in this
parameter.

For example, we know from lattice simulations that
the number of degrees of freedom is somewhat less than
the value corresponding to
a gas of free gluons even at very high temperatures, where we expect a
deconfined gluon plasma. For a review with updated results, see
 \cite{pasechnik}.
 Another factor is that, even though we consider a regime where
we expect the energy to be given by the exchange of massive gluons,
in principle, one can also have some contribution from glueballs.
Generally, such contributions will be of the form
$\sum_\alpha C_\alpha \, e^{ - 2 m_\alpha x}$, where
$C_\alpha$ are some constants and $m_\alpha$ is the mass of the
glueball labeled by $\alpha$. (The factor of 2 in the exponent is because we need two transits between the plates to complete a loop for the virtual
particle, as explained in \cite{KN1}. It is also clear from the worldline method of calculating the Casimir energy.)
While a proper calculation
of such terms will require information about how glueballs can couple to the plates and what boundary conditions should be imposed on the glueball fields,
we can estimate the importance of such terms by comparison with glueball
masses, which are typically higher than the value of the dynamical mass $m$ 
for the gluons. So each such contribution is exponentially small compared to
our formulae (\ref{13}) and (\ref{4d-16}), even though there could be many such terms.

The lowest lying glueball in 2+1 dimensions
has a mass $4.716$ for $SU(2)$
and $4.330$ for $SU(3)$ (in units of $\sqrt{\sigma_{\rm F}}$), while the corresponding values
in 3+1 dimensions are $3.781$ and
$3.405$ \cite{AthT}.
In 2+1 dimensions, the values for the dynamical mass of the gluon are
$0.921 \sqrt{\sigma_{\rm F}}$ for $SU(2)$ and
$0.846 \sqrt{\sigma_{\rm F}}$ for $SU(3)$, as seen from (\ref{12b}).
Since the glueball mass is much higher, we might expect that
any contribution from the glueballs will be very small
in 2+1 dimensions, so that we can expect good
agreement with (\ref{13}).
In 3+1 dimensions, the mass for the gluon (for $SU(3)$) was 
in the range $1$ to $1.41$ in units of $\sqrt{\sigma_{\rm F}}$.
The glueball masses are closer to this value compared to 2+1 dimensions,
so it is possible that there is some extra contribution to
the energy, beyond (\ref{4d-16}).
This might explain why a better fit is obtained in the four-dimensional
case if we use a value for $C$ not equal to $1$. Even so, we think it should not
deviate too much from $1$.
Once again, an unambiguous set of lattice data 
is needed to reach any definite conclusion.

Finally, we comment on the use of the formula
(\ref{com2}).
If we set
$M_{\rm Cas} = 0$, formula (\ref{com1}) (with $\nu = 2$)
correctly reduces to the
Casimir energy due to one massless field (for each 
direction in the Lie algebra), in agreement with the expectation
of one physical polarization for the gauge field in 2+1 dimensions.
However, even though setting 
$M_{\rm Cas} = 0$ in (\ref{com2}) (with $\nu = 3$)
also gives the correct form of the energy for ${\rm dim}G$ massless fields in
3+1 dimensions, it still corresponds to only one polarization, not two, as
expected. 
This may be a point of concern regarding the use of the formula
(\ref{com2}) to obtain a fit to the lattice data.

\section{Conclusions}

We have calculated the Casimir energy for the standard arrangement of two parallel plates for nonabelian gauge fields in 2+1 and 3+1 dimensions. In both cases, the result corresponds to that of ${\rm dim}G$ massive scalar fields, 
with the appropriate number of polarizations (one in 2+1 and two in 3+1 dimensions). 
These modes are the gauge-invariant versions of the gluons
with their propagators having a dynamically generated mass.
We also compared the analytic formulae,
with the available results based on lattice
simulations, all for the case of chromoelectric (PEC) boundary conditions.
The agreement in 2+1 dimensions is excellent. We emphasize that
our formula in 2+1 dimensions has no adjustable parameters.
The two quantities needed, namely the string tension and the
propagator mass of the gluon, both follow from our Hamiltonian
analysis and are numbers calculated purely based on group theory.
It is important that we do have an excellent agreement, 
for both $SU(2)$ and $SU(3)$ for which lattice data is available,
without any adjustable parameters.

In 3+1 dimensions, the propagator mass is a parameter which defines the
theory. It is the parameter of dimensional transmutation, the analog of
$\Lambda_{\rm QCD}$ in our gauge-invariant analysis.
Here too the agreement with lattice data is good, but there seem to be
differences in the lattice data from two different groups, so
an unambiguous conclusion is not possible at this stage.
\vskip .05in
\noindent\underline{\it Note added}:\\
As this paper was being typed up, a paper discussing 
analytic calculations of
the Casimir energy was posted on the arXiv \cite{dudal}.
The authors consider the Curci-Ferrari (CF) model which 
describes a Yang-Mills theory with massive gluons
in 2+1 and 3+1 dimensions.
There is soft breaking of the BRST (Becchi-Rouet-Stora-Tyutin)
symmetry due to the mass term,
but it is argued that the CF model is an acceptable description
for YM theory with dynamical generation of mass.
Formulae similar to our equations (\ref{12a}) and
(\ref{4d-16}) are obtained and comparison with 
lattice data is done with the prefactor $C$ and $m$ 
as fitting parameters.
We mention a couple of points worth emphasizing in this regard.
Our analysis in both 2+1 and 3+1 dimensions are fully gauge-invariant.
The mass term, even in 3+1 dimensions, is gauge invariant,
although the truncation of the theory to the quadratic terms of the action
is all that is effective for the Casimir energy.
Further, as mentioned above, we do not have adjustable parameters in
2+1 dimensions.

\bigskip

This work was supported in part by the U.S. National Science Foundation Grants No. PHY-2412479 and PHY-2412480.

\section*{Appendix: Covariantizing the theory on $\mathbb{CP}^2$}
\def\theequation{A\arabic{equation}}
\setcounter{equation}{0}
In section 4 we have seen that by considering the 4d nonabelian gauge theory on $\mathbb{CP}^2 =SU(3)/U(2)$ we can use the completeness of the Wigner $\D$-functions on $SU(3)$ to find a general parametrization of the gauge fields as in (\ref{4d-1}). The complex gauge fields on $\mathbb{CP}^2$ are parametrized in terms of matrix variables ($M, M^{\dag}, \chi, \chi^{\dag}$) on which gauge transformations act homogeneously ($M\rightarrow M^g = gM$), allowing for a straightforward separation of gauge degrees of freedom. For a more detailed account see \cite{KMN}.

While complexification in 4d proves useful for a manifestly gauge-invariant description, the resulting theory is ultimately a theory in complex space rather than in $\mathbb{R}^4$. Imposing a complex structure in 4d Euclidean space breaks the full rotational symmetry of Euclidean space (reducing the rotational group from $O(4)$ to $U(2)$). We can restore this symmetry by integrating over all the possible complex structures in $\mathbb{R}^4$. To this end, we employ a twistor space construction, where the complex projective space $\mathbb{CP}^3$ can be viewed as a bundle of local complex structures over $S^4$, i.e., locally, $\mathbb{CP}^3 \sim S^4 \times \mathbb{CP}^1$, where $\mathbb{CP}^1$ parametrizes the space of local complex structures. 
In particular, we will link our gauge theory in the large volume limit of $\mathbb{CP}^2$ to a theory on $S^4$ with a particular complex structure given by a point on $\mathbb{CP}^1$. Integrating over the $\mathbb{CP}^1$ fiber---equivalent to integrating over all possible complex structures---restores Euclidean covariance.

An explicit realization of $\mathbb{CP}^3 \sim S^4 \times \mathbb{CP}^1$ is given in twistor theory, where the homogeneous coordinates on $\mathbb{CP}^3$, $Z = (Z_1, Z_2, Z_3, Z_4)$, with the identification $Z\sim \lambda Z$, $\lambda \in \mathbb{C} - \{0\}$, can be written in terms of a pair of two-component spinors $w$ and $\pi$ in the following way
\beq
Z= (w_1,w_2,\pi_1,\pi_2)
\label{a1}
\eeq
Here, $\pi=(\pi_1,\pi_2)\neq (0,0)$ are homogenous coordinates on $\mathbb{CP}^1$, defined up to an overall scaling $\pi \sim \lambda \pi$, with $\lambda \in \mathbb{C}-\{0\}$. The coordinates on $S^4$, $x^\mu$, $\mu =1,2,3,4$, are introduced via the incidence relations---relations linking spacetime to twistor space:
\beq
w = (x^4 -i \vec{\sigma} \cdot \vec{x} ) \pi
\label{a2}
\eeq
where $w=(w_1,w_2)$ and $\vec{\sigma} = (\sigma_1, \sigma_2, \sigma_3)$ are the Pauli matrices.

The basic idea is to take $w$ and $\bar{w}$ as coordinatizing our theory 
on $\mathbb{CP}^2$ in the large radius limit, effectively
$\mathbb{CP}^2 \rightarrow \mathbb{C}^2$.
\footnote{For us the fields will vanish at infinity,
and the difference between the large radius limits of
$\mathbb{CP}^2$, $S^4$ and $\mathbb{R}^4$ will not be
important. 
Thus $w$ and $\bar{w}$ can be taken to coordinatize $\mathbb{C}^2$,
or $S^4$ (with the one point removed), when we take the radius
to be large.} A point on the $\mathbb{CP}^1$ fiber corresponds to a particular complex combination of the real coordinates $x^\mu$---for example, $\pi = (0,i)$ yields $w_1 = x^1 - i x^2$, $w_2 = -x^3 +ix^4$. Integrating over the $\mathbb{CP}^1$ coordinates will restore Euclidean covariance. 

Following the incidence relations in (\ref{a2}), it is straightforward to relate the complex gauge fields, $A_i$, $\bA_{\bi}$, $i,\bi = 1,2$, to the real ones, $A_\mu$, $\mu=1,2,3,4$, as follows
\beqar
A_i &=& {1 \over 2 \bar{\pi} \cdot \pi} ( \bar{\pi}_i A_4(x) + i (\bar{\pi}\sigma_a)_i A_a(x) ) \nonumber\\
\bA_{\bi} &=& {1 \over 2 \bar{\pi} \cdot \pi} (\pi_i \, A_4(x) -i (\sigma_a \pi)_i \, A_a(x) )
\label{a3}
\eeqar
where $a=1,2,3$. 
We can further parametrize the real gauge fields as follows
\beq
A_4 = -i (\del_4 \theta + \del_a \vf_a), \hskip.2in A_a = -i (\del_a \theta - \del_4 \vf_a - \e_{abc} \del_b \vf_c )
\label{a4}
\eeq
where the gauge fields $A_\mu$ are chosen to be anti-Hermitian, $A_\mu^{\dag} = - A_\mu$, for Hermitian fields $\theta^{\dag} = \theta$, $ \vf^{\dag} = \vf$. 

Since the complex gauge fields, $A_i, \bA_{\bi}$, are in general functions of both $x^\mu$ and $\mathbb{CP}^1$ coordinates, we expect the same to be the case for the matrix parameters $M, M^{\dag}, \chi, \chi^{\dag}$. 
From (\ref{a4}) and the parametrization in (\ref{4d-1}), the complex matrix parameters to first order in the real fields $\theta, \vf$ are given by
\beqar
M &=& 1 + i \theta + {\bar{\pi} \sigma_a \pi \over \bar{\pi} \cdot \pi} \vf_a + \cdots \nonumber\\
\chi &=& {\e^{ij} \bar{\pi}_i (\bar{\pi} \sigma_a)_j \over \bar{\pi} \cdot \pi}  \vf_a + \cdots
\label{a5}
\eeqar
where the elipses indicate terms with more powers of $\theta$ and $\vf$.

We now have everything we need to covariantize the action of the complex theory on $\mathbb{C}^2$. As detailed in \cite{KMN} the Jacobian determinant for the change of variables $A_i, \bA_{\bi} \rightarrow H, \chi, \chi^{\dag}$ contributes an effective action of the form
\beq
\Gamma = C S_{\rm wzw}(H) - \mu_{\rm ren}^2 \int \Tr \ba \cdot H a H^{-1} + \Gamma_{\log \e} +\cdots
\label{a6}
\eeq
where the elipsis includes terms of higher scaling dimension that are subdominant in the low energy regime. The coefficient $C$ is a finite coefficient and $\mu^2_{\rm ren}$ is the finite renormalized value of mass for the $a_i, \ba_{\bi}$ fields. The log-divergent terms, $\Gamma_{\log\e}$, are in general complicated and take the form
\beqar
\Gamma_{\log \e} &=& {\log \e\over 12} \int \Tr \left[ (\bnabla(\nabla H H^{-1}))^2 + (\ba H a H^{-1})^2 \right. \nonumber\\
&& \hskip .8in + g^{i\bi} g^{j \bar{j}}[\ba_{\bi}, Ha_i H^{-1}] \bnabla_{\bar{j}}(\nabla_j H H^{-1}) \nonumber\\
&&\hskip .8in \left. - g^{i\bi} g^{j \bar{j}} \left(\bnabla_{\bi}(\nabla_j H H^{-1}) [\ba_{\bar{j}},Ha_iH^{-1}]+ \bnabla_{\bi} \ba_{\bar{j}} \D_i (Ha_jH^{-1}) \right) \right]
\label{a7}
\eeqar
where $\D$ is a covariant derivative with $-\nabla H H^{-1}$ as its connection, and $\e$ is the UV cutoff. The log divergence needs to be renormalized through the introduction of counterterms.

Overall, the effective action for the Yang-Mills theory in terms of the complex matrix parameters is given by
\beq
\S = - \Gamma + S_{\rm YM}
\label{a8}
\eeq
where $S_{\rm YM} = - {1 \over 2g^2} \int F_{\mu\nu}^2$ is the standard Yang-Mills action.

To covariantize this action we first reexpress the complex matrix parameters in terms of real fields through (\ref{a5}), and then we integrate over the $\mathbb{CP}^1$ variables. To show this by example, let us covariantize the WZW action up to quadratic order in $\vf$ fields. First, we write the WZW action on $\mathbb{CP}^3$ in the following way
\beqar
S_{\rm wzw} (H) &=& {1\over 2\pi} \int_{\mathbb{CP}^1} d\mu(\pi,\bar{\pi}) \int {d^2 w d^2 \bar{w} \over (\bar{\pi}\cdot \pi)^2}\, \Tr \,\del H \cdot \bdel H^{-1} \nonumber\\
&& -{1\over 2\pi} \int_{\mathbb{CP}^1, w}\int dt g^{i\bi} \Tr H^{-1} \del_t H [H^{-1} \del_iH, H^{-1} \bdel_{\bi} H]
\label{a9}
\eeqar
where $d\mu(\pi,\bar{\pi})$ is the volume element on $\mathbb{CP}^1$. To quadratic order in $\vf$'s only the first term is relevant. Rewriting $H$ in terms of $\vf$ fields
\beqar
S_{\rm wzw} &=& -{1\over 2\pi} \int_{\mathbb{CP}^1} {\bar{\pi} \sigma_a \pi\over \bar{\pi}\cdot \pi} {\bar{\pi} \sigma_b \pi \over \bar{\pi} \cdot \pi} \int d^4x \Tr \left(\del_\mu \vf_a \del_\mu \vf_b\right) + \theta(\vf^3) \nonumber\\
&=& -{1\over 6\pi} \int d^4x \, \Tr\left( \del_\mu \vf_a \del_\mu \vf_a\right) + \theta(\vf^3)
\label{a10}
\eeqar
Integrating over the $\mathbb{CP}^1$ coordinates in the other terms in $\S$ leads to
\beq
\S = {1 \over 2 g^2}\int d^4x \left[ (\Box \vf_a^\a)^2 + m^2 (\del \vf^\a_a)^2 \right] + \theta (\vf^3)
\label{a11}
\eeq
where $g$ is the renormalized coupling constant, and $m^2 = {c_A g^2 \over 3} (C/\pi +\mu^2_{\rm ren})$ with $c_A$ being the value of the quadratic Casimir operator in the adjoint representation of the gauge group. $\vf_a^\a$ are the Lie algebra components of the fields $\vf_a = t_\a \vf_a^\a$. 

Upon covariantization, the mass term for the $a_i$ fields combines with the WZW term producing a single mass term for the real fields $\vf$. Up to quadratic order in $\vf$, $\Gamma_{\log\e}$ has the same form as $S_{\rm YM}$, thus it simply contributes to the renormalization of $g^2$. We can redefine our fields as $\vf \rightarrow g \vf$ resulting in
\beq
\S = {1\over 2} \int \left[ (\Box \vf^\a_a)^2 + m^2 (\del \vf^\a_a)^2 \right]
\label{a12}
\eeq

In this Appendix, we have used $\alpha$ for the color index, 
$a, b, c$ for the spatial vector components, so as to avoid confusion
with $i$, $\bi$ used for the complex components and coordinates.
To be in conformity with the text, we will now switch to using $a$
as the color index, and $i, j, k$ for the spatial vector indices.
The action is thus
\beq
\S = {1\over 2} \int \left[ (\Box \vf^a_i)^2 + m^2 (\del \vf^a_i)^2 \right]
\label{a13}
\eeq
This is the action (\ref{4d-4}) used in the text.



\begin{thebibliography}{99}

\bibitem{chern1} M. N. Chernodub, V. A. Goy, A. V. Molochkov, Ha Huu Nguyen,
Phys. Rev. Lett. {\bf 121}, 191601 (2018).

\bibitem{chern2} M.N. Chernodub, V. A. Goy, A. V. Molochkov, 
A.S. Tanashkin, \PR~{\bf D 108} 1, 014515 (2023) [arXiv:2302.00376[hep-lat]].

\bibitem{ngwenya} B.A. Ngwenya, arXiv:2505.0287[hep-lat];
B.A. Ngwenya, A.K. Rothkopf and W.A. Horowitz,
arXiv:2507.21333[hep-lat].

\bibitem{KKN1} D.~Karabali and V.~P.~Nair,
  Nucl.\ Phys.\  B {\bf 464}, 135 (1996)
  [arXiv:hep-th/9510157];
  Phys.\ Lett.\ B {\bf 379}, 141 (1996)
  [hep-th/9602155];
  D.~Karabali, C.~j.~Kim and V.~P.~Nair,
  Nucl.\ Phys.\  B {\bf 524}, 661 (1998)
  [arXiv:hep-th/9705087];
  D.~Karabali, C.~j.~Kim and V.~P.~Nair,
  Phys.\ Lett.\  B {\bf 434}, 103 (1998)
  [arXiv:hep-th/9804132].

\bibitem{KNY} D.~Karabali, V.~P.~Nair and A.~Yelnikov,
  Nucl.\ Phys.\ B {\bf 824}, 387 (2010)
  [arXiv:0906.0783 [hep-th]].

\bibitem{Ham2+1} For  recent review, see
V.P. Nair, Int. J. Mod. Phys. {\bf A 38}, 33n34, 2330016 (2023)
[arXiv:2308.13926[hep-th]].

\bibitem{KN1} D. Karabali and V.P. Nair,  
Phys. Rev. {\bf D 98}, 105009 (2018).

\bibitem{GPY} D. Gross, R. Pisarski and L. Yaffe, \RMP~ {\bf 53}, 43 (1981) and references therein.

\bibitem{magmass}
V.P. Nair, \PL~ {\bf B 352}, 117 (1995);
G. Alexanian and V.P. Nair, \PL~{\bf B 352}, 435 (1995);
W. Buchm\"uller and O. Philipsen, Nucl.Phys. B443 (1995) 47;
W. Buchm\"uller and O. Philipsen, \PL~{\bf B397}, 112 (1997);
R. Jackiw and S.Y. Pi, \PL~{\bf B368}, 131 (1996);
{\it ibid} {\bf B403}, 297 (1997);
V.P. Nair, Rev. in Math. Phys. {\bf 33}, 214002 (2021)
[arXiv:1910.06051[hep-th]].\\
The possibility of calculating the magnetic mass by summing Feynman diagrams was addressed in J.M. Cornwall, \PR~{\bf D26}, 1453 (1982). The paper J.M. Cornwall, W-S. Hou, J.E. King, \PL~{\bf B153}, 173 (1985) gave the bound $m/e^2 \geq 0.58$; the later paper J.M. Cornwall, \PR~{\bf D57}, 3694 (1998) gave a modified procedure with the estimate $0.248 \approx 0.25$, as
quoted. See also
 J.M. Cornwall, Phys. Rev. {\bf D10}, 500 (1974); 
{\it ibid.} {\bf D57}, 3694 (1998);
J. M. Cornwall and B. Yan, \PR~{\bf D 53}, 4638 (1996);
J.M. Cornwall, \PR~{\bf D57}, 3694 (1998);
J.M. Cornwall, \PR~{\bf D76}, 025012 (2007).

\bibitem{KMN} D. Karabali, A. Maj and V.P. Nair,
\PR~{\bf D 106} 8, 085012 (2022) [arXiv:2110.11926[hep-th]];
\PR~ {\bf D 106} 8, 085013 (2022) [arXiv:2206.09985[hep-th]].

\bibitem{casimir} For general reviews on the Casimir effect, see
K.A. Milton, {\it Recent developments in Casimir effect}, 
J. Phys. Conf. Ser. {\bf 161}, 012001 (2009);
K. A. Milton, {\it The Casimir Effect: Physical Manifestations of Zero-Point Energy} (World Scientific, 2001);
M. Bordag, U. Mohideen and V.M. Mostepanenko, Phys. Rept. {\bf 353}, 1 (2001);
M. Bordag, G.L. Klimchitskaya, U. Mohideen and V.M. Mostepanenko, {\it Advances in the Casimir Effect}, International Series of Monographs on Physics, Vol. 145
(Oxford University Press, 2009),  pp.1-768.

\bibitem{KN2} D. Karabali and V.P. Nair, \PR~{\bf D 87}, 105021 (2013);
T.R. Govindarajan and V.P. Nair, \PR~{\bf D89}, 025020 (2014);
D. Karabali and V.P. Nair, \PR~{\bf D 92}, 125003 (2015).

\bibitem{asorey} M. Asorey, C. Iuliano and F. Ezquerro, 
Physics 2024, 6(2), 613;
M. Asorey, F. Ezquerro and M. Pardina, Ann. Phys.
481, 170195 (2025) [arXiv:2501.18072[hep-th]].

\bibitem{teper} M. Teper, \PR~ {\bf D59}, 014512 (1999);
B. Lucini and M. Teper, \PR~ {\bf D66}, 097502 (2002);
H.B. Meyer and M.J. Teper, \NP~{\bf B668}, 111 (2003);
N.D. Hari Dass and P. Majumdar,
\PL ~{\bf B 658}, 273 (2008)  [arXiv:hep-lat/0702019];
J. Kiskis and R. Narayanan, 
\JHEP~ 0809:080 (2008) [arXiv:0807.1315[hep-th]];
B.H. Wellegehausen, A. Wipf and C. Wozar,
\PR~ {\bf D 83}, 016001 (2011)  [arXiv:1006.2305[hep-lat]].

\bibitem{teper2} B.~Bringoltz and M.~Teper, Phys.\ Lett.\  B {\bf 645}, 383 (2007)  [arXiv:hep-th/0611286].

\bibitem{LMY}
R.~G.~Leigh, D.~Minic and A.~Yelnikov,
  \PRL~{\bf 96}, 222001 (2006);
  \PR ~{\bf D76}, 065018 (2007).

\bibitem{KNrobust}   D. Karabali and V.P. Nair, \PR~{\bf D 77}, 025014
(2008) [arXiv:0705.2898[hep-th]].

\bibitem{AN} A.~Agarwal and V.~P.~Nair,
  Nucl.\ Phys.\ B {\bf 816}, 117 (2009)
  [arXiv:0807.2131 [hep-th]].
  
 \bibitem{mass-scalar}
M.V. Cougo-Pinto, C. Farina, A.J. Segui-Santonja,
Lett. Math. Phys. {\bf 31}, 309 (1994);
S. Mobassem, Mod. Phys. Lett. {\bf A29}, 1450160 (2014)
(arXiv:1403.0501[quant-ph]).

\bibitem{saharian} A.A. Saharian and H.H. Asatryan, arXiv:2507.07267[hep-th].

\bibitem{dudal} D. Dudal, P. de Fabritiis and S. Stouten, 
arXiv:2509.07256[hep-th].

\bibitem{MIT-piston} A. Edery and V. Marachevsky, \PR~{\bf D 78},  025021 (2008).

\bibitem{dudal1} D. Dudal, P. Pais and L. Rosa, \PR~{bf D 102}, 016026 (2020). 

\bibitem{LucT} B. Lucini and M. Teper, \PR~{\bf D 66}, 097502 (2002).
 
 \bibitem{masslat} I. L. Bogolubsky, E. M. Ilgenfritz, M. Muller-Preussker, and A. Sternbeck,
\PL~ {\bf B 676} (2009) 69;
P. O. Bowman, et al.,  \PR~ {\bf D 76} (2007)
094505;
  O. Oliveira and P. J. Silva,
Proc. Sci. LAT2009 ({2009}) 226 [arXiv: 0910.2897];
A. Cucchieri and T. Mendes,Proc. Sci., LAT2007 ({2007}) 297 [arXiv: 0710.0412];
Phys. Rev. {\bf D81}, 016005 (2010);A. Cucchieri, D. Dudal, T. Mendes and N. Vandersickel, \PR~ {\bf D 85} (2012) 094513;
A.G. Duarte, O. Oliveira and P.J. Silva, \PR~{\bf D 94}, 014502 (2016).


\bibitem{mass2} For reviews on the subject, see:
J.M. Cornwall, J. Papavassiliou and D. Binosi, {\it The Pinch Technique and its Application
to Nonabelian Gauge Theories}, Cambridge University Press (2011);
N. Vandersickel and D. Zwanziger, ``The Gribov problem and QCD dynamics", {\it Phys. Rep.} {\bf 520} (2012) 175;
A.C. Aguilar, D. Binosi and J. Papavassiliou, ``The gluon mass generation mechanism: A concise primer",
{\it Front. Phys.} {\bf 11} (2016) 111203.

\bibitem{papavassiliou}
M.N. Ferreira, J. Papavassiliou, J.M. Pawlowski and
N. Wink, arXiv:2508.20568 [hep-ph] and references therein.

\bibitem{pasechnik} R. Pasechnik and M. Sumbera,
Universe {\bf 3}, 7 (2017)   [arXiv:1611.01533[hep-ph]].

\bibitem{AthT} A. Athenodorou and M. Teper, \JHEP~{\bf 2021}, 82
(2021).







\end{thebibliography}
\end{document}